\documentclass[aps,prd,12pt,a4paper,groupedaddress,preprintnumbers,floatfix,nofootinbib]{revtex4}
\usepackage[english]{babel}
\usepackage{amsmath}
\usepackage{graphicx}
\usepackage{color}
\usepackage{amssymb}
\usepackage{hyperref}
\usepackage{dcolumn}
\usepackage{bm}
\usepackage[margin=2cm]{geometry}
\begin{document}
\preprint{DCP-12-02}

\title{Three generations of Higgses and the cyclic groups}

\author{
Alfredo Aranda,$^{1,2}$\footnote{Electronic address:fefo@ucol.mx}
Cesar Bonilla,$^{3}$\footnote{Electronic address:rasec.cmbd@gmail.com}
J. Lorenzo D\'{\i}az-Cruz,$^{2,3}$ \footnote{Electronic address:jldiaz@fcfm.buap.mx} }

\affiliation{$^1$Facultad de Ciencias - CUICBAS,\\
  Universidad de Colima, Colima, M\'exico\\
  $^2$Dual C-P Institute of High Energy Physics, M\'exico \\
  $^3$C.A. de Part\'{\i}culas, Campos y Relatividad,
  FCFM-BUAP, Puebla, Pue., Mexico}

\date{\today}

\begin{abstract}
Multi Higgs doublet models are interesting extensions of the Standard Model that can be related to flavor. The reason is that most flavor models usually involve the presence of several additional scalar fields. In this work we present an analysis that shows that for renormalizable flavor models based on the cyclic group of order $N$, if there is one flavored SU(2) double Higgs per generation, the smallest $N$ that can be used to reproduce the Nearest-Neighbor-Interaction texture for the quark mass matrices is $N=5$. Results for the Higgs spectrum and consistency under $K - \bar{K}$ mixing in a specific model with $Z_5$ are also presented.
\end{abstract}
\maketitle

The Yukawa matrices of the Standard Model (SM) parametrize our ignorance on the possible relations and perhaps origin of the mass spectrum and mixing angles of all fundamental fermions. Given the experimental determination of fermion masses (so far only mass squared differences in the case of neutrinos), the entries of the Cabibbo-Kobayashi-Maskawa (CKM) matrix (including its CP violating phase), and the values of the neutrino mixing angles encoded into the Pontecorvo-Maki-Nakagawa-Sakata (PMNS) matrix, it is possible to determine, in a model independent way, certain {\it textures} for the fermion mass matrices (obtained from the Yukawa matrices) that upon diagonalization reproduce those results. In particular, considering for the moment the quark sector, the so-called Nearest-Neighbor-Interaction (NNI) texture~\cite{Branco:1988iq,FXZ} successfully reproduces the quark masses and CKM angles and phase. The NNI texture has the general form
\begin{eqnarray} \label{NNI}
M^{u,d}=\left( \begin{array}{ccc}
0          &      \star       &  0   \\
\star      &        0         & \star \\
0          &      \star       & \star
\end{array} \right),
\end{eqnarray}
and can always be obtained from general mass matrices through a proper choice of flavor basis in the SM~\cite{Branco:1988iq,FXZ}. The lepton sector can be included by either assuming a diagonal mass matrix for the charged leptons and diagonalizing the neutrino mass matrix 
with the PMNS matrix, or by considering the possibility that the PMNS matrix receives contributions from both sectors. Extending the SM in order to incorporate descriptions of fermion masses and mixing angles thus necessarily requires either to predict or accommodate those specific textures.

Intimately related to these issues is the scalar sector of the SM and its extensions. The spontaneous breaking of Electroweak (EW) symmetry in the SM, through the non-zero vacuum expectation value (vev) of a Higgs field, is what generates the mass matrices from the Yukawa matrices. The SM incorporates this mechanism in a minimal fashion, i.e. by postulating the existence of one doublet (under SU(2)) scalar field. Other choices and possibilities have been explored and there is a vast literature associated to them. Among the most popular extensions are those that consider the case (within many different settings) with two Higgs doublets, generically called two Higgs doublet models (THDM - see~\cite{Branco:2011iw} for a recent review). Higgs triplets (in combination with doublets) have also been studied extensively~\cite{Grimus:2004hf,Gunion:1989ci,DiazCruz:2007tf}. In most cases these extensions are motivated by the scalar sector phenomenology and do not necessarily add much to the problem described above. Most cases 
involve flavor-blind Higgs fields that do not directly participate into the structure of the mass matrices.

An interesting scenario where the Higgs fields necessarily participate in the flavor structure is the one of renormalizable flavor models~\cite{renormalizable,AZ4}. In this type of models a horizontal flavor symmetry, continuous or discrete, is added to the SM gauge group symmetry in such a way as to reproduce the observed mass and mixing angle patterns by only using renormalizable Lagrangians. This requirement has two immediate and interesting consequences: i) there must be more than one SU(2) doublet scalar and ii) at least some of them must transform non-trivially under the flavor symmetry. 

In this letter we consider the case of renormalizable models involving three Higgs SU(2) doublets ($H_1$, $H_2$, $H_3$), the SM gauge and fermion particle content, and a discrete Abelian flavor symmetry $Z_N$.  The purpose is to find the smallest $Z_N$ that can be used to obtain the NNI textures in Eq.(\ref{NNI}) in a setting with three Higgses (the smallest realization of the NNI structure with two Higgs doublets requires $Z_4$~\cite{Branco:2010tx}). This setting is interesting from the point of view of minimal extensions to the SM that could give useful hints to more elaborated and ambitious extensions. Studying the case of three SU(2) doublets is motivated by the fact that most flavor models, renormalizable and non-renormalizable, usually require the presence of several additional scalar fields. Our purpose here is to show that having one flavored Higgs {\it per generation}, can lead to interesting possibilities.

In order accomplish our goal, the following considerations are taken into account: 
\begin{itemize}
\item[i)] Each non-zero entry in the mass matrices contains contributions from a single vev. The condition is required by the fact that if several Higgses contribute to the same entry, they necessarily will have the same $Z_N$ charge, making them virtually indistinguishable.
\item[ii)] One of the Higgses will be assigned neutral charge while the other two will be related by conjugation. This is necessary in order to use the same three Higgses in both the up and down quark sectors (in the case of even $N$ it is also possible to consider one of the Higgses to have charge $N/2$ instead of the neutral charge, however this requires larger groups than the neutral case. See Appendix~\ref{appendix1}).
\item[iii)] One of the Higgses contributes exclusively to the $3-3$ entry of the mass matrix for the up-type quark. This is motivated by the fact that the top quark is the heaviest and resembles the familiar case of models based non-Abelian flavor groups where the fermions are put in ${\bf 2} \oplus {\bf 1}$ representations~\citep{Aranda:2000tm}.   
\end{itemize}

Given these considerations, the possible textures for the up-type quarks have the general form ($k \neq l \neq m$)
\begin{eqnarray}
\label{allcases} \nonumber
M^{u}_{A1} &\sim &\left( \begin{array}{ccc}
0          &      v_k       &  0   \\
v_k      &        0         & v_l \\
0          &      v_l       & v_m
\end{array} \right), \ \
M^{u}_{A2} \sim \left( \begin{array}{ccc}
0          &      v_k       &  0   \\
v_l      &        0         & v_k \\
0          &      v_l       & v_m
\end{array} \right), \ \
M^{u}_{A3} \sim \left( \begin{array}{ccc}
0          &      v_k       &  0   \\
v_l      &        0         & v_l \\
0          &      v_k       & v_m
\end{array} \right),
 \\ \nonumber \\
M^{u}_{B1} &\sim& \left( \begin{array}{ccc}
0          &      v_k       &  0   \\
v_k      &        0         & v_k \\
0          &      v_l       & v_m
\end{array} \right), \ \
M^{u}_{B2} \sim \left( \begin{array}{ccc}
0          &      v_k       &  0   \\
v_l      &        0         & v_l \\
0          &      v_l       & v_m
\end{array} \right),
\end{eqnarray}
where $v_{(k,l,m)}$, $(k,l,m)=1,2,3$ denote the Higgs vevs.

The charge assignments for the fermion and scalar fields is parametrized as 
\begin{eqnarray}
\label{charges}
\overline{Q} & \simeq & (q_1, q_2, q_3), \ 
U_R \simeq (u_1, u_2, u_3), \
D_R \simeq (d_1, d_2, d_3), \\
\tilde{{\cal H}} &\equiv&  (\tilde{H_1}, \tilde{H_2}, \tilde{H_3}) \simeq (h_1, h_2, h_3) \ ,
\end{eqnarray}
where $q_i, u_i, d_i, h_i \in Z_N$, $\overline{Q}$ is the left-handed quark SU(2) doublet, $U_R$ ($D_R$) is the up-type (down-type) right handed quark SU(2) singlet, and $\tilde{H_i} = i\sigma_2 H^*$. The $Z_N$ charges of the bilinears in the Yukawa terms of the Lagrangian can be represented by 
\begin{eqnarray}
\label{generalyukawa}
{\cal Y}^u_{ij} = q_i+u_j \ mod(N). 
\end{eqnarray}
We now present the results showing the complete analysis for one case.

{\it Case A1:} Without loss of generality, let $k=1$, $l=2$, and $m=3$ for $M_{A1}^u$ in Eq.~(\ref{allcases}). From condition ii) above we have the following three possibilities for the Higgses charges: a) ($h_1=0, h_2=a,h_3=-a$), b) ($h_1=a, h_2=0,h_3=-a$), and c) ($h_1=-a, h_2=a,h_3=0$) with $a \in Z_N$. 

Case a) leads to the following constraints on the fermion bilinear charges Eq.~(\ref{generalyukawa}):
\begin{eqnarray}
\label{A1aconstraints1}
{\cal Y}^u_{12} = 0, \ {\cal Y}^u_{21} = 0, \ {\cal Y}^u_{23} = -a, \  {\cal Y}^u_{32} = -a, \ {\cal Y}^u_{33} = a, 
\end{eqnarray}    
and
\begin{eqnarray}
\label{A1aconstraints2}
{\cal Y}^u_{11}, \ {\cal Y}^u_{13}, \ {\cal Y}^u_{22}, \ {\cal Y}^u_{31} \neq  (0,-a,a).
\end{eqnarray}    

Using Eq.~(\ref{A1aconstraints1}) we find that the fermion assignments become
\begin{eqnarray}
\label{A1a-assignments}
q_1 & = & -c, \ \ q_2 =  -c -3a, \ \ q_3  = -c -a, \\
u_1 & = & c+3a, \ \ u_2 =  c, \ \ u_3  = c+2a, \\
d_1 & = & c+3a, \ \ d_2 =  c, \ \ d_3  = c+2a,
\end{eqnarray}
where $c,a \in Z_N$. The last step is to satisfy the relations in Eq.~(\ref{A1aconstraints2}) which for this case become:
\begin{eqnarray}
\label{lastcondition}
(3a, 2a, -3a) \neq \left\{\begin{array}{c} 0 \\ -a \\ a \end{array} \right. \ mod(N) \ .
\end{eqnarray}

Since $2a=0$ in $Z_2$ and $3a=0$ in $Z_3$, these two groups are discarded. For $Z_4$, $a \in (0,1,2,3)$. Since $a \neq 0$, then  $a \in (1,2,3)$. Since $2a \neq 0$, then $a \in (1,3)$. If $a=1$, then $3a = 3 = -1 = -a$ and thus $a \neq 1$. $Z_4$ must be discarded. For $Z_5$ $a \in (0,1,2,3,4)$ and again, since  $a \neq 0$, then  $a \in (1,2,3,4)$. If $a=1$, then $-a=4$, $2a=2$, $3a=3$, $-3a=2$, and thus all conditions in Eq.~(\ref{lastcondition}) are satisfied and the smallest group that works in this case is $Z_5$.

A similar analysis for all other possibilities shows that 
\begin{itemize}
\item $Z_5$ is the smallest group that can be used in this setting and corresponds to the case A1-a). Larger groups can also be used in this case.
\item $Z_6$ is the smallest possibility for case A1-c). Larger groups can also be used in this case.
\item All other cases do not satisfy the necessary conditions for any $Z_N$.
\end{itemize}

Thus, the smallest group that can be used within this setting is $Z_5$. Note that the analysis includes the 
down-type quark sector as well by choosing the same charge assignments as those of the up-type. 
However, it is also possible to obtain the required texture with other assignments. 
This is of course a model dependent issue. As for leptons, it is straightforward to 
generalize to the charged leptons by giving them the same charge assignments used for 
the down-type quarks. Neutrino masses and mixing can then be included either considering 
radiative mass generation~\cite{Zee:1980,Babu:1988qv,Fukugita:2003en}, non-renormalizable terms, 
or seesaw through the introduction of right-handed neutrinos~\cite{seesaw}. This is a model dependent 
question and we do not explore it further in this letter except to mention that for instance, 
in a model like the one in case A1-a), the lepton sector can reproduce exactly the one presented in~\cite{AZ4}. 
A specific model and its phenomenology will be presented in a future publication.

We note that, given the condition on the scalar charge assignments, 
it is possible to write a general $Z_N$ invariant potential (a general discussion on Abelian symmetries 
in multi-Higgs models can be found in~\cite{Ivanov:2011ae}). Since one of the Higsses is required to 
be neutral and the remaining two to be conjugate of each other, and for clarity, let us use the 
following notation: denote by $H$ the neutral one and by $\Phi_a$, $a=1,2$ the remaining two. 
The potential is then given by
\begin{eqnarray}
\label{potential} \nonumber
V(H,\Phi_a) &=& \mu_0^2 |H|^2 + \mu_a^2|\Phi_a|^2 + \mu_{0a}^2\left( \Phi_a^{\dagger}H + h.c \right)
+\mu_{12}^2 \left(\Phi_1^{\dagger}\Phi_2 + h.c \right)
+ \lambda_0 \left( |H|^2\right)^2 \nonumber \\ 
&+&\lambda_a \left( |\Phi_a|^2 \right)^2 
+ \lambda_{0a} |H|^2|\Phi_a|^2 + \lambda_{12} |\Phi_1|^2|\Phi_2|^2 +
\tilde{\lambda}_{ab}|\Phi^{\dagger}_a \tilde{\Phi}_b|^2 +\lambda'_{0a}\Phi_{a}^{\dagger}HH^{\dagger}\Phi_{a}\notag\\
&+&\lambda_{3}\left(\Phi_{1}^{\dagger}H\Phi_{2}^{\dagger}H + h.c\right),
\end{eqnarray}
where the terms proportional to $\mu_{0a}$ and $\mu_{12}$ are $Z_N$ soft breaking terms 
required in order to obtain the correct electromagnetic invariant vacuum. Denoting the scalars by
\begin{eqnarray} \label{scalars}
H = \left( \begin{array}{c}
H^+ \\
\frac{1}{\sqrt{2}}\left( v_0 + h+ i A_0\right)\end{array}\right) , \ \
\Phi_a = \left( \begin{array}{c}
\Phi_a^+ \\
\frac{1}{\sqrt{2}}\left( v_a + \phi_a + i A_a\right)\end{array}\right) ,
\end{eqnarray}
where $v_0$ and $v_a$ denote the vevs of $H$ and $\Phi_a$ respectively, the minimization conditions become
\begin{eqnarray}
\label{minimization}
\mu_0^2 &=& -\frac{\lambda'_{01}v_{1}^{2}v_{0}+2\lambda_{3}v_{1}v_{2}v_{0}+\lambda'_{02}v_{2}^{2}v_{0}+\lambda_{01} v_1^2 v_0  + \lambda_{02} v_2^2 v_0 + 2 \lambda_0 v_0^3 + 2 v_1 \mu_{01}^2 + 2 v_2 \mu_{02}^2}{2 v_0} , \\
\mu_1^2 &=& -\frac{\lambda_{01}v_{1}v_{0}^{2}+\lambda_{3}v_{2}v_{0}^{2}+ 2\lambda_{1} v_1^3 +\lambda_{12}v_1 v_2^2 + \lambda_{01} v_1 v_0^2 +2 v_2 \mu_{12}^2 +2 v_0 \mu_{01}^2}{2v_1} , \\
\mu_2^2 &=& -\frac{\lambda_{02}v_{2}v_{0}^{2}+\lambda_{3}v_{1}v_{0}^{2}+\lambda_{12} v_1^2 v_2 +2\lambda_{2}v_2^3 + \lambda_{02}v_2 v_0^2 +2v_1 \mu_{12}^2 + 2v_0 \mu_{02}^2}{2v_2} .
\end{eqnarray}

In order to perform a numerical analysis for the Higgs mass spectrum it is necessary to study
a specific model. Taking as an example the case A1-a) for $Z_5$ we have performed a scan over the parameter space of the model. We find that there are large regions of parameter space consistent with current experimental values and bounds. As an example we present in Table~\ref{table} one particular set of parameters (a complete study including a statistical analysis involving a $\chi^2$ fit, possible collider signatures, as well as a compelte analysis of the lepton sector, is under preparation and will be published in a future paper) that gives the following spectrum: For the three CP-even scalar masses we obtain 
(in GeV): $125.7$, $700.9$, and $892.1$; for the two massive CP-odd scalars we obtain (in GeV): $670.4$ and $894.1$; and for the two charged scalars (again in GeV): $678.7$ and $895.3$.
\begin{center}
\begin{table} [h]
\begin{tabular}{ ||c|c|c|c|c|c||}
  \hline                        
$v_{1}$~(GeV) & $v_{2}$~(GeV) & $v_{3}$~(GeV)& $\mu_{12}^2$~(GeV)$^2$ & $\mu_{01}^2$~(GeV)$^2$ & $\mu_{02}^2$~(GeV)$^2$ \\
 \hline
$210$ & $69.7$ & $107.5$ & $ -(350)^2$ & $-(400)^2$ & $-(450)^2$ \\
  \hline  
\end{tabular}

\begin{tabular}{ ||c|c|c|c|c|c|c|c|c|c|c|c|| }
  \hline                        
 $\lambda_{0}$&$\lambda_{1}$&$\lambda_{2}$&$\lambda_{12}$&$\lambda_{01}$&$\lambda_{02}$&$\lambda'_{01}$ &$\lambda'_{02}$&$\lambda_{3}$\\
  \hline
 0.63927&-0.561199&0.160189&0.0779788&-0.758485&0.426743&-0.543321&-0.582515&-0.0203623\\
  \hline  
\end{tabular}
\caption{One set of parameter values consistent with current experimental data.}
\label{table}
\end{table}
\end{center}

Furthermore, since the
model contributes to flavor changing transitions at tree level, we also incorporate the constraint coming from $K - \bar{K}$ mixing. This is done by computing $\Delta M_{K}$ from the effective Hamiltonian \cite{Buras:1998raa,Buras:2001ra,DdPl},
\begin{equation}
\mathcal{H}^{\Delta S=2}_{eff}=\frac{G_{F}^{2}M_{W}^{2}}{16\pi^{2}}\sum_{i}C_{i}(\mu)Q(\mu) 
\end{equation}
where $G_{F}=1.16639\times10^{-5} \text{GeV}^{-2}$ is the Fermi constant and $C_{i}$ are the Wilson coefficients. 
Consequently, the $K-\overline{K}$ mixing is governed by the neutral scalar interactions with the first and second 
down type quark families, then
\begin{align}
&Q_{2}^{LR}=(\bar{s}P_{L}d)(\bar{s}P_{R}d)\\
&Q_{1}^{SLL}=(\bar{s}P_{L}d)(\bar{s}P_{L}d)\\
&Q_{1}^{SRR}=(\bar{s}P_{R}d)(\bar{s}P_{R}d)
\end{align}
whose coefficients, at leading order, are
\begin{align}
&C_{2}^{LR}=-\frac{16 \pi^{2}}{G_{F}^{2}M_{W}^{2}}\left(\frac{m_{d}m_{s}}{v_{1}^{2}}\right)\sum_{a=1}^{3}\frac{U_{2a}U_{1a}}{m_{h_{a}^{0}}}\\
&C_{1}^{SLL}=-\frac{16 \pi^{2}}{G_{F}^{2}M_{W}^{2}}\left(\frac{m_{d}m_{s}}{v_{1}^{2}}\right)\sum_{a=1}^{3}\frac{U_{2a}^{2}}{m_{h_{a}^{0}}}\\
&C_{1}^{SRR}=-\frac{16 \pi^{2}}{G_{F}^{2}M_{W}^{2}}\left(\frac{m_{d}m_{s}}{v_{1}^{2}}\right)\sum_{a=1}^{3}\frac{U_{1a}^{2}}{m_{h_{a}^{0}}} 
\end{align}
where all fields are in the mass basis and $U_{ab}$ denotes the scalar mixing matrix. The $K-\overline{K}$ mixing is given by the off-diagonal term in the neutral 
$K$-meson mass matrix 
\begin{equation}
M_{12}^{K}\equiv\frac{\Delta M_{K}}{M_{K}}=7.2948\times10^{-15}.
\end{equation}
where $\Delta M_{K}$ is given by~\cite{DdPl}
\begin{equation}\label{DeltaMK}
\Delta M_{K}=2\text{Re}\langle\overline{K^{0}}|\mathcal{H}^{\Delta S=2}_{eff}|K^{0}\rangle= 
\frac{G_{F}^{2}M_{W}^{2}}{12\pi^{2}}M_{K}F_{K}^{2}\eta_{2}\hat{B}_{K}\left[\bar{P}_{2}^{LR}C_{2}^{LR}+
\bar{P}_{1}^{SLL}\left(C_{1}^{SLL}+C_{1}^{SRR}\right)\right],
\end{equation}
where $F_{K}=160$~MeV is the K-meson decay constant, $M_{K}=497.6$~MeV is the K-meson mass, and $\hat{B}_K$ and $\eta_2$ include the QCD running effects. 
We follow the notation in~\cite{DdPl} so that $\bar{P}_{1}^{SLL}=-9.3$, $\bar{P}_{2}^{LR}=30.6$, 
$\eta_{2}=0.57$, $\hat{B}_{K}=0.85\pm 0.15$. Consequently, we obtain
\begin{equation}
M_{12}^{K}=\frac{4 }{3}F_{K}^{2}\eta_{2}\hat{B}_{K}(m_{d}m_{s})\underbrace{\frac{1}{v_{1}^{2}}
\sum_{a=1}^{3}\left[\bar{P}_{2}^{LR}\frac{U_{2a}U_{1a}}{{m_{h_{a}^{0}}^{2}}}+\bar{P}_{1}^{SLL}
\left(\frac{U_{2a}^{2}}{{m_{h_{a}^{0}}^{2}}}+\frac{U_{1a}^{2}}{{m_{h_{a}^{0}}^{2}}}\right)\right]}_{\equiv\mathcal{F}_{C}}.
\end{equation}
Taking into account that $m_{d}\sim 5$~MeV and $m_{s}\sim 100$~MeV, it is possible 
to establish that the constraint coming from $KK$ mixing is satisfied in our model when
\begin{equation}
\mathcal{F}_{C}<\frac{3M_{12}^{K}}{4F_{K}^{2}\eta_{2}\hat{B}_{K}m_{d}m_{s}}\simeq 8.82 \times 10^{-10}(\text{GeV}^{-4}).
\end{equation}
Taking the values in Table~\ref{table} we obtain
\begin{equation}
 \mathcal{F}_{C}^{Z_{5}}=5.5025\times10^{-11} \text{GeV}^{-4}.
\end{equation}
 
In conclusion we have shown that the smallest cyclic group that can be used to generate the NNI textures for the 
quark mass matrices, in the context of three SU(2) Higgs doublets where one couples only to the $3-3$ entry of the 
up-type quark mass matrix, is $Z_5$. We have outlined the analysis that led to this result and have presented the 
general scalar potential for such a scenario. The results presented here can be of use to model builders interested 
in flavor models with three Higgs doublets.


\begin{acknowledgments}
This work was supported in part by CONACYT and SNI (Mexico). C.B. thanks the CUICBAS of Universidad de Colima for its hospitality while part of this work was carried out.
\end{acknowledgments}

\begin{appendix}

\section{Analysis for even $N$}
\label{appendix1}

Recall from Eq.~\eqref{allcases} all possible textures for the up-type quarks:
\begin{eqnarray}
\label{allcases2} \nonumber
M^{u}_{A1} &\sim &\left( \begin{array}{ccc}
0          &      v_k       &  0   \\
v_k      &        0         & v_l \\
0          &      v_l       & v_m
\end{array} \right), \ \
M^{u}_{A2} \sim \left( \begin{array}{ccc}
0          &      v_k       &  0   \\
v_l      &        0         & v_k \\
0          &      v_l       & v_m
\end{array} \right), \ \
M^{u}_{A3} \sim \left( \begin{array}{ccc}
0          &      v_k       &  0   \\
v_l      &        0         & v_l \\
0          &      v_k       & v_m
\end{array} \right),
 \\ \nonumber \\
M^{u}_{B1} &\sim& \left( \begin{array}{ccc}
0          &      v_k       &  0   \\
v_k      &        0         & v_k \\
0          &      v_l       & v_m
\end{array} \right), \ \
M^{u}_{B2} \sim \left( \begin{array}{ccc}
0          &      v_k       &  0   \\
v_l      &        0         & v_l \\
0          &      v_l       & v_m
\end{array} \right),
\end{eqnarray}
where $v_{(k,l,m)}$, $(k,l,m)=1,2,3$ denote the Higgs vevs. 

In general the Higgses have charges $\tilde{H}=(h_1=a,h_2=b,h_3=c)$ where $a,b,c\in Z_{N}$. Now lets consider case $A_1$ 
\begin{itemize}
 \item[$A_{1}$)] The fermion fields have the following $Z_{N}$ charges:
 \begin{align}
 &q_{1}=\alpha, \ \ q_{2}=-c+\alpha-a+2b,\ \ q_{3}=\alpha-a+b\notag\\
 &u_{1}=2a+c-\alpha-2b,\ \ u_{2}=a-\alpha,\ \ u_{3}=a+3b+c-\alpha
 \end{align}
 and the constraints become
\begin{eqnarray}
\label{cond}
(\mathcal{Y}_{11}, \mathcal{Y}_{22}, \mathcal{Y}_{13},\mathcal{Y}_{31}) \neq \left\{\begin{array}{c} a \\ b \\ c \end{array} \right. \ mod(N) \ .
\end{eqnarray}
where,
\begin{align}
&\mathcal{Y}_{11}=2a+c-2b,\ \ \mathcal{Y}_{22}=-c+2b,\ \ \mathcal{Y}_{13}=a+3b+c\ \ \text{and}\ \ \mathcal{Y}_{31}=a-b+c.
\end{align}

Taking into consideration the condition that two of the Higgses are related by conjugation, we have the following possibilities, a) $\tilde{H}=(d,e,-e)$, b) $\tilde{H}=(e,d,-e)$ and c) $\tilde{H}=(-e,e,d)$ where, $e,d\in Z_{N}$. From the additional condition that the remaining Higgs is either neutral or, for $N$ even, $N/2$, we have the two cases $d=0$ for any $N$ and $d=N/2$ for $N$ even. The case $d=0$ has been presented in the paper and we now present the analysis for $d=N/2$. In this case the constraints for each possibility of $A_1$ are:
\begin{itemize}
\item[a)]
\begin{align}
 &\mathcal{Y}_{11}=N-3e,\ \ \mathcal{Y}_{22}=3e,\ \ \mathcal{Y}_{13}=N/2+2e\ \ \text{and}\ \ \mathcal{Y}_{31}=N/2-2e,
\end{align}
which are satisfied with an Abelian symmetry group of order $N\geq8$. 
\item[b)]
\begin{align}
 &\mathcal{Y}_{11}=e-N,\ \ \mathcal{Y}_{22}=e+N,\ \ \mathcal{Y}_{13}=3N/2\ \ \text{and}\ \ \mathcal{Y}_{31}=-N/2,
\end{align}
which can not be satisfied with any Abelian symmetry.
\item[c)]
\begin{align}
 &\mathcal{Y}_{11}=-4e+N/2,\ \ \mathcal{Y}_{22}=-N/2+2e,\ \ \mathcal{Y}_{13}=N/2+2e\ \ \text{and}\ \ \mathcal{Y}_{31}=N/2-2e,
\end{align}
where the minimal Abelian symmetry group is again $Z_{8}$.
\end{itemize}
\item The constraints obtained from cases $A_{2}$, $A_{3}$, $B_{1}$, and $B_{2}$  cannot be satisfied for any $N$.
\end{itemize}
\end{appendix}

\end{document}